\documentclass[iop,revtex4]{emulateapj}

\usepackage{epsfig,graphicx}
\usepackage{amsmath}
\usepackage{natbib}
\usepackage{times}
\usepackage{txfonts}
\usepackage{hyperref}
\usepackage{microtype}
\shorttitle{The s-Process Enrichment of the Globular Clusters M4 and M22}
\shortauthors{Shingles et al.}

\providecommand{\noopsort}[1]{}
\begin{document}

\title{The s-Process Enrichment of the Globular Clusters M4 and M22}

\author{Luke J. Shingles\altaffilmark{1}$^\dag$, Amanda I. Karakas\altaffilmark{1}\altaffilmark{2}, Raphael Hirschi\altaffilmark{2}\altaffilmark{3}, Cherie K. Fishlock\altaffilmark{1}, David Yong\altaffilmark{1}, Gary S. Da Costa\altaffilmark{1}, Anna F. Marino\altaffilmark{1}}

\altaffiltext{1}{Research School of Astronomy and Astrophysics, Australian National University, Canberra, ACT 2611, Australia}
\altaffiltext{2}{Institute for the Physics and Mathematics of the Universe (WPI), University of Tokyo, 5-1-5 Kashiwanoha, 277-8583 Kashiwa, Japan}
\altaffiltext{3}{Astrophysics Group, EPSAM, Keele University, Lennard-Jones Labs, Keele, ST5 5BG, UK}

\altaffiltext{\dag}{Email: luke.shingles@anu.edu.au}

\begin{abstract}
We investigate the enrichment in elements produced by the slow neutron-capture process ($s$-process) in the globular clusters M4 (NGC 6121) and M22 (NGC 6656). Stars in M4 have homogeneous abundances of Fe and neutron-capture elements, but the entire cluster is enhanced in $s$-process elements (Sr, Y, Ba, Pb) relative to other clusters with a similar metallicity. In M22, two stellar groups exhibit different abundances of Fe and $s$-process elements. By subtracting the mean abundances of $s$-poor from $s$-rich stars, we derive $s$-process residuals or empirical $s$-process distributions for M4 and M22. We find that the $s$-process distribution in M22 is more weighted toward the heavy $s$-peak (Ba, La, Ce) and Pb than M4, which has been enriched mostly with light $s$-peak elements (Sr, Y, Zr). We construct simple chemical evolution models using yields from massive star models that include rotation, which dramatically increases $s$-process production at low metallicity. We show that our massive star models with rotation rates of up to 50\% of the critical (break-up) velocity and changes to the preferred $^{17}$O($\alpha$,$\gamma$)$^{21}$Ne rate produce insufficient heavy $s$-elements and Pb to match the empirical distributions. For models that incorporate AGB yields, we find that intermediate-mass yields (with a $^{22}$Ne neutron source) alone do not reproduce the light-to-heavy $s$-element ratios for M4 and M22, and that a small contribution from models with a $^{13}$C pocket is required. With our assumption that $^{13}$C pockets form for initial masses below a transition range between 3.0 and 3.5 M$_\sun$, we match the light-to-heavy s-element ratio in the s-process residual of M22 and predict a minimum enrichment timescale of between 240 and 360 Myr. Our predicted value is consistent with the 300 Myr upper limit age difference between the two groups derived from isochrone fitting.
\end{abstract}


\section{Introduction}\label{sec:intro}
\setcounter{footnote}{3}
The assumption that globular clusters (GCs) are simple stellar populations (i.e., populations of stars that formed simultaneously from gas of a uniform chemical composition) has made them ideal laboratories for the study of low-mass stellar evolution \citep{Moehler:2001gc} and enabled their ages to be accurately determined. This has aided cosmology by setting a lower limit on the age of the universe \citep{Chaboyer:1996hh,Dotter:2010ek}. However, the simple stellar population model of GCs has been undermined by spectroscopic studies that reveal significant star-to-star abundance variations ($\gtrsim 1$ dex) in the light elements from C to Al \citep[e.g.,][]{Cottrell:1981ha,Carretta:2009kr,Denissenkov:2014bc}. Similar variations found in unevolved stars show that the chemical variations were initially present in the star-forming gas rather than being the result of nucleosynthesis and mixing within the observed stars \citep{Cannon:1998ki,Gratton:2001ew}. More recently, photometric studies have independently confirmed the existence of multiple populations in the form of split main sequences and sub-giant branches in color-magnitude diagrams \citep[e.g.,][]{Piotto:2007br,Piotto:2009ds,Milone:2008dz}.

The light element patterns that exist almost exclusively in globular clusters \citep[i.e., rarely in field stars and open clusters, see][]{Gratton:2000wg,DeSilva:2009ki} include anti-correlations between the abundances of C and N, Na and O, and sometimes Mg and Al, typically with a C+N+O abundance that is constant within observational errors. The abundance patterns depict a H-burning process at high temperature ($>80$ MK\footnote{1 MK = $10^6$ K.}) combined with dilution by varying amounts of unprocessed material, although the stellar sites where this burning takes place and the mechanism of dilution are presently not well understood \citep{Denisenkov:1990vz,Langer:1993dx,Decressin:2007bf,Prantzos:2007bs,DOrazi:2010kb,DErcole:2011jc}.

In contrast to the light elements which vary in abundance, GCs are typically homogenous in [Fe/H]\footnote{We use the standard spectroscopic notation $[A/B] = \log(A/B) - \log(A/B)_\sun$, where A and B are abundances by number and $\odot$ denotes the solar abundance.} \citep[$\sigma$\textless 0.05 dex,][]{Carretta:2009di} and in the abundances of neutron-capture elements \citep[$Z> 30$;][]{Gratton:2004dy,Yong:2006ia,Yong:2008in,DOrazi:2010fs}. Exceptions are known, including $\omega$ Centauri \citep{Norris:1995kc,Smith:2000cb,Johnson:2010fs}, M22 \citep{Marino:2009je}, NGC 1851 \citep{Yong:2008ja,Villanova:2010bi,Carretta:2011eo}, M2 \citep{Yong:2014wd}, M15 \citep{Sneden:1997fp,Sobeck:2011dv}, and possibly NGC 2419 \citep{Cohen:2012cd}.

Neutron-capture elements refer to elements with atomic number Z $>30$, because production of these elements is almost entirely by a process of neutron captures and $\beta^-$-decay reactions. Depending on whether the average neutron-capture rates are less than or greater than the average rate of $\beta^-$-decay reactions, the processes are divided into the slow ($s$-process) and rapid ($r$-process) neutron-capture processes \citep{Burbidge:1957hf}. Although most heavy elements can be synthesized by both processes, elements whose production is dominated by the $r$- or the $s$-process in solar system material are commonly referred to as $r$- and $s$-process elements.

Due to the large uncertainties involved in numerically modeling nucleosynthesis by the $r$-process, the $r$-only component of a heavy element distribution is often inferred from solar system material by subtracting the $s$-process component, which itself may be determined either theoretically \citep[e.g.,][]{Arlandini:1999eh,Goriely:1999wj,Sneden:2008cf} or empirically \citep[e.g.,][]{Simmerer:2004ib}.

The $s$-process takes place at low neutron densities \citep[$\leq 10^{14}$ cm$^{-3}$;][]{Busso:1999ig} and operates exclusively on nuclides that are very close to stability, as nuclei that become unstable following neutron capture have time to $\beta$-decay back to stability before additional neutron captures occur. In the build-up of progressively heavier elements via the $s$-process, bottlenecks form around nuclides with `magic' numbers of neutrons (e.g., 50, 82, 126) which form nuclear structures that are more stable against neutron capture than their neighbors \citep{Busso:1999ig}. Three major peaks develop: a light $s$-peak (Sr, Y, Zr), a heavy $s$-peak (Ba, La, Ce), and a peak at Pb, with the light peak forming first and the heavier peaks forming later with increasing neutron exposure.

Globular clusters provide laboratories to test and explore our understanding of stellar nucleosynthesis. One cluster that has been studied extensively is M22 (NGC 6656), which exhibits internal variation in [Fe/H] and $s$-process abundances that are bimodally distributed and neatly separate into two groups \citep{Marino:2009je,Marino:2011kz,DaCosta:2011jh}. While there are other well-studied clusters with Fe and $s$-process variation (e.g., $\omega$ Centauri), the simpler chemical evolution history of M22 relative to more complex systems like $\omega$ Centauri makes it an attractive system for testing theories about $s$-process variation in globular clusters more generally.

Even among globular clusters that are homogenous in their abundances of Fe and neutron-capture elements there exist puzzles surrounding the chemical evolution of the $s$-process elements. For example, M4 is a fairly typical mono-metallic metal-poor GC ([Fe/H] $=-1.18$; \citealt{Carretta:2009di}), except that it has super-solar abundances of $s$-process peak elements \citep[e.g., Rb, Y, Zr, La, Ba, Pb;][]{Brown:1992dw,Ivans:1999hf}. The origin of the $s$-process elements in M4 and M22 is speculated on in the literature, but often on the basis of individual stellar yields \citep[e.g.,][]{Roederer:2011hw} rather than a full investigation using a chemical evolution model. Very recently, \citet{Straniero:2014jk} presented the first comparison of the $s$-process distributions of M4 and M22 with the summed contribution from a generation of AGB stars at the metallicity of M22 ([Fe/H]$=-1.8$).

In this paper, we present simplified chemical evolution models of the heavy elements in globular clusters and predict the abundance variations that arise from $s$-process production by 1) massive stars with rotation, or 2) a generation of AGB stars that span a range of stellar masses. We then compare our chemical abundance predictions with the observed abundances of stars in M4 and M22. The success or failure of the individual models gives us insight into the stellar sites and timescales of $s$-process enrichment in globular clusters, as well as highlighting the shortcomings of current stellar nucleosynthesis models.

\section{The s-process in massive stars}
We define as massive stars those with sufficient mass to eventually form a collapsing core of Fe and end their lives as core-collapse supernovae. Current estimates for the lower-limit of initial mass required to meet this condition are around 8 to 12 M$_\sun$, with lower masses required at lower metallicities \citep{Langer:2012jy,Nomoto:2013js,Jones:2013iw}.

In massive stars, neutron-capture nucleosynthesis takes place during pre-supernova evolution and possibly also during the supernova itself. During convective core He-burning and shell He- and C-burning, neutrons are released via the $^{22}$Ne($\alpha$,n)$^{25}$Mg reaction \citep{Peters:1968bf,Raiteri:1992dn,Meyer:1994br,The:2007gb}.

The production of $^{22}$Ne occurs via He-burning of $^{14}$N left over from H-burning in the CNO cycle. In models without rotation $^{22}$Ne is secondary since its yield depends on the initial amount present plus any formed from $\alpha$-capture onto $^{14}$N, which itself is limited by the initial abundance of C+N+O. Hence, there is very little $s$-process production at low metallicity in non-rotating models. Some production of heavy elements in massive stars does take place (the weak $s$-process) but this is mainly concentrated around elements of the first $s$-peak near Y, with virtually no heavy $s$-elements or Pb being produced \citep{Beer:1992jv,Pignatari:2010ir}.

In models of massive stars that do include rotation, rotationally-induced mixing transports primary $^{12}$C and $^{16}$O produced in the convective He-core to the H-burning shell, where it is then converted into $^{14}$N via the CN-cycle \citep{Meynet:2006bh}. The primary $^{14}$N is then mixed into and burnt in the He core, resulting in an almost-primary production of $^{22}$Ne \citep{Hirschi:2007br} that dramatically increases $s$-process yields at low metallicity.

\begin{figure}
 \begin{center}\includegraphics[width=1.0\columnwidth]{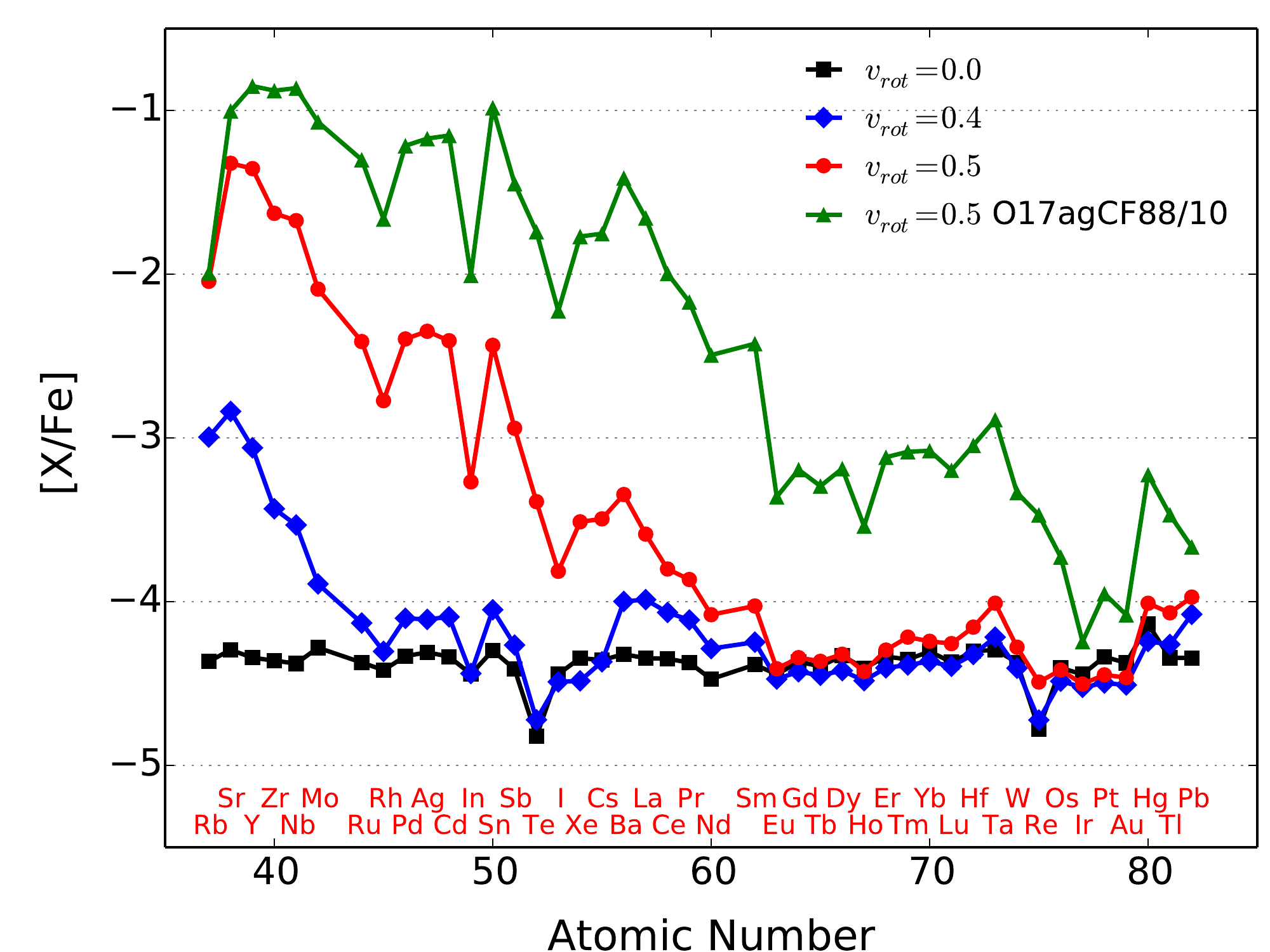}\end{center}
 \caption{Abundance ratios with Fe relative to the solar values in the pre-supernova yields of 25 M$_\odot$ models at [Fe/H]$=-3.8$ with several initial rotation rates. The rotation rate is given in units of the critical velocity ($v_{crit}$). Yields from \citet{Frischknecht:2012il} with zero-metallicity explosive Fe yields from \citet{Limongi:2012ei}.}
 \label{fig:rmsyields}
\end{figure}

\citet{Pignatari:2008ec} present the first $s$-process yields for a rotating massive star with their 25 M$_\sun$ model. They find that rotation increases $s$-process yields by orders of magnitude and alters the standard weak $s$-process distribution with a peak of production between Sr and Ba. The high production of heavy $s$-elements in their model is due to the use of the very low \citet{Descouvemont:1993kc} rate for the $^{17}$O($\alpha$,$\gamma$)$^{21}$Ne reaction, which is disfavored by recent experiments. \citet{Frischknecht:2012il} present $s$-process yields from a set of massive models with updated reaction rates and find that rotation leads to the complete consumption of Fe-seeds at metallicities below $Z=10^{-3}$ and an increase to the production of elements near the Ba peak at the expense of the Sr peak as metallicity decreases.

Figure \ref{fig:rmsyields} presents the heavy-element yields of 25 M$_\odot$ massive star models with several initial rotation rates from \citet{Frischknecht:2012il}. This figure demonstrates that under the condition of fast rotation, the $s$-process production in massive stars at low metallicity begins to include elements that would otherwise be associated uniquely with AGB stars (e.g., Ba, La, and Pb). For this reason, massive rotating stars must be considered as a possible source of the neutron-capture elements in globular clusters.

In this study we use the pre-supernova yields of neutron-capture elements calculated from a grid of rotating massive stars (including those used to generate Figure \ref{fig:rmsyields}) with initial masses of 15, 20, 25, and 40 M$_\sun$ at initial metallicities of $Z=10^{-5}$ ([Fe/H]$=-3.8$) and $Z=10^{-3}$ ([Fe/H]$=-1.8$) with $\alpha$-enhanced initial compositions as described in \citet{Frishknecht:2012vi}. The rotation rates of the models are specified by their initial velocity at the equator as a fraction of the break-up velocity ($v_{crit}$, the velocity at which centrifugal force balances gravity).

For elements $Z \le 26$, we use the zero-metallicity explosive yields of \citet{Limongi:2012ei}. Although supernova yields presently carry large uncertainties, the effect of varying the Fe yields will be to scale our resulting heavy element distributions up and down while leaving the ratios between elements unchanged. The supernova shockwave will not significantly affect the $s$-process production and hence the $s$-process yields are approximated by their pre-supernova values although the mass yields and to a lesser extent the s-process distribution, will depend on the mass cut \citep{Tur:2009ji}.

\section{The $s$-process in AGB stars}\label{sec:agbsprocess}

In low to intermediate mass (0.8 to 8 M$_\sun$) stars, the $s$-process takes place during the thermally-pulsing AGB phase of evolution. For further details on AGB stellar evolution and nucleosynthesis, we refer to the reviews by \citet{Herwig:2005jn} and \citet{Karakas:2014tp}.

\begin{figure}
 \begin{center}\includegraphics[width=1.0\columnwidth]{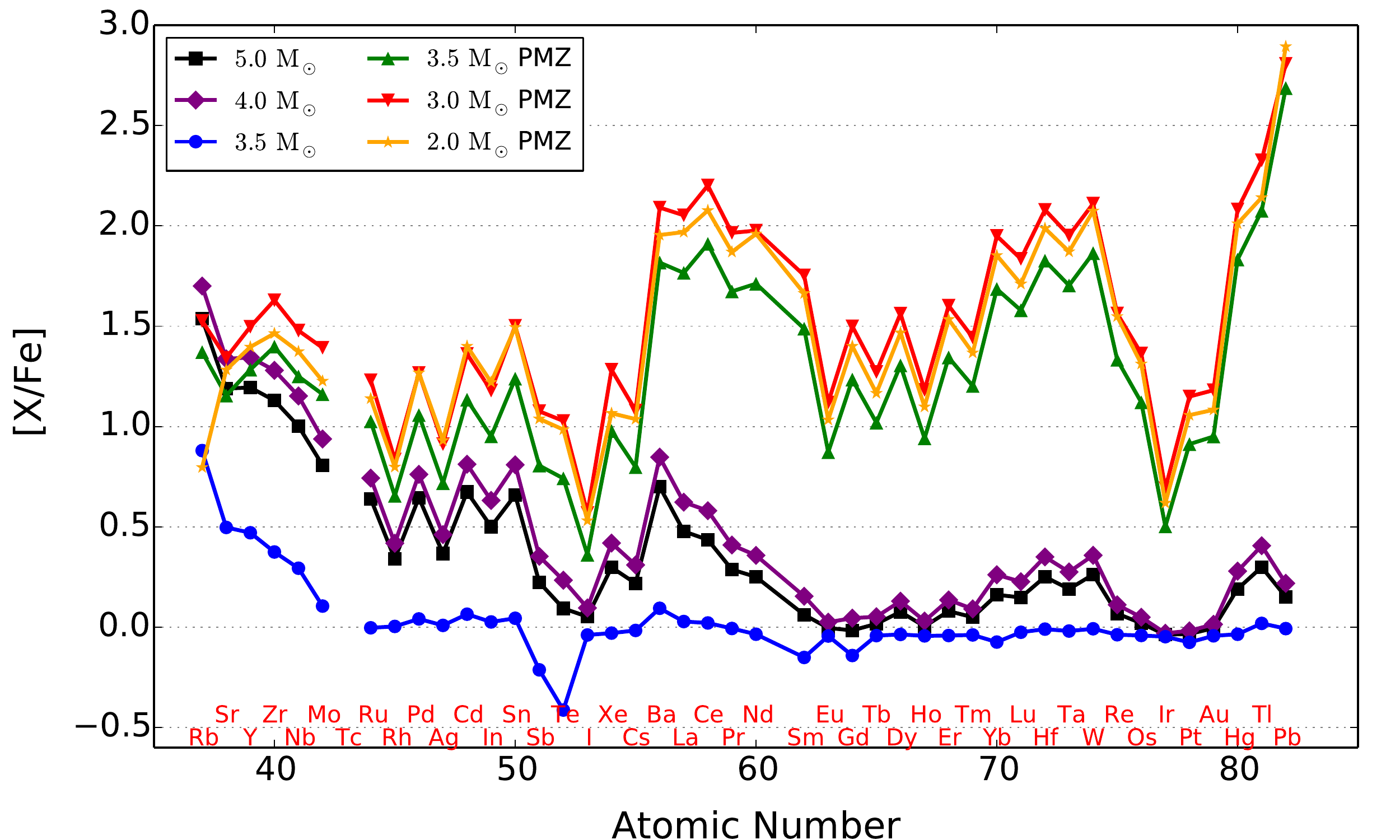}\end{center}
 \caption{Abundance ratios with Fe relative to their solar values in the yields of AGB models at [Fe/H]$=-1.2$ with several different initial masses. Models labelled `PMZ' include a partial mixing zone. Yields from \citet{Fishlock:OGIWnrqE}.}
 \label{fig:agbyields}
\end{figure}

Figure \ref{fig:agbyields} shows the average composition of the stellar ejecta of AGB models selected from the full grid which includes masses of 2.5, 2.75, 3.00, 3.25, 3.5, 4.0, 4.5, 5.0, 6.0, and 7.0 M$_\sun$ at a metallicity of $Z=0.001$ ([Fe/H]$=-1.2$ scaled solar), and are taken from \citet{Fishlock:OGIWnrqE}. This figure displays the transition between the $s$-process yields from low-mass stars ($\lesssim$ 4 M$_\sun$) to intermediate-mass stars as a result of the $^{22}$Ne neutron source becoming active. This transition mass also roughly coincides with our assumed upper limit initial masses for $^{13}$C pockets in AGB stars of 3 or 3.5 M$_\odot$ at [Fe/H] $=-1.2$. For the rest of this section, we briefly summarize the operation of the $s$-process in AGB stars.

With increasing initial mass, the maximum temperature in the intershell obtained during a thermal pulse also increases. A consequence is that in stars $\lesssim$ 4 M$_\sun$, fewer thermal pulses are accompanied by a substantial activation of the $^{22}$Ne($\alpha$,n)$^{25}$Mg reaction. Instead, free neutrons for the $s$-process are mainly released by radiative $^{13}$C-burning via the $^{13}$C($\alpha$,n)$^{16}$O reaction, which is active at temperatures as low as 90 MK \citep{Cameron:1955eb,Straniero:1995ed}. Producing the required $^{13}$C has been a challenge for stellar modelers, as the $^{13}$C abundance left behind by the H-burning shell is too low to allow for sufficient $s$-processing, and the convective region following a thermal pulse cannot extend into the H-rich region above the $^{12}$C-rich intershell \citep{Iben:1975js}.

Current AGB models achieve s-process nucleosynthesis via a $^{13}$C pocket in the following manner: protons from the envelope are `partially mixed' beyond the formal convective border into the $^{12}$C-rich intershell region, thus enabling the CN cycle reactions $^{12}$C(p,$\gamma$)$^{13}$N($\beta^+$)$^{13}$C \citep{Gallino:1998eg,Arlandini:1999eh}. The mixing process is required to have only marginal efficiency, otherwise the newly-synthesized $^{13}$C is readily destroyed by further proton captures to make $^{14}$N, which is a neutron poison, i.e., its large neutron-capture cross section makes it an efficient absorber of free neutrons. For models that include a partial-mixing zone (PMZ) by inserting an exponential profile of protons below the inner edge of the envelope convective zone, a $^{13}$C pocket is formed below a pocket of $^{14}$N \citep{Cristallo:2009kn,Lugaro:2012ht}. In our post-process AGB nucleosynthesis models (including those used to calculate the yields shown in Figure \ref{fig:agbyields}) we include a PMZ of $2\times 10^{-3}$ M$_\sun$ at the deepest extent of each third dredge-up episode for initial masses $\leq 3.0$ M$_\sun$ and a $1\times 10^{-3}$ M$_\sun$ PMZ for selected models at 3.25 and 3.5 M$_\sun$. The technique we use to include a PMZ is identical to \citet{Lugaro:2012ht} and we refer the reader to that paper for more details. We discuss the uncertainties related to $^{13}$C pockets in Section \ref{sec:modeluncertainties}.

Figure \ref{fig:agbyields} illustrates that low-mass stars produce significant quantities of heavy $s$-peak elements and Pb at low metallicity and confirms previous results in the literature \citep{Travaglio:2001en,VanEck:2001cj,VanEck:2003el,Lugaro:2012ht}. This is because the neutron source $^{13}$C is primary (independent of metallicity) while at low metallicity fewer Fe-seed nuclei (the most abundant heavy element) are available \citep{Clayton:1988tf}. With a large neutron supply per Fe-seed, neutrons are preferentially captured by heavier nuclei and the abundance distribution is shifted toward higher atomic numbers.

Figure \ref{fig:agbyields} also shows that the yields of elements heavier than Sr are significantly lower in models with masses $\geq 4$ M$_\sun$ at a metallicity of [Fe/H]$=-1.2$. This is because the dominant neutron source in these models is the $^{22}$Ne($\alpha$,n)$^{25}$Mg reaction, which is active at temperatures above about 300 MK \citep{Cameron:1960hj,Iben:1975bn,Goriely:2000vs}. The high temperatures and $^{22}$Ne nuclei required to activate this source are found near the base of the He intershell in convective zones driven by He-shell flashes. Thus, neutrons are briefly exposed to a relatively large number of Fe seeds at the base of the flash-driven convective zone, and the resulting $s$-process distribution in intermediate-mass stars is mostly weighted towards the light $s$-peak near Sr-Y-Zr, with lower yields of heavy $s$-elements compared to lower mass stars.

In summary, the change from the $^{22}$Ne source operating in convective pulses to radiative $^{13}$C-burning during the interpulse phase creates a dramatic change in the distribution of heavy elements between models above and below the transition mass of around 3-4 M$_\sun$. The precise mass of this transition is dependent on the choice of the highest mass to include a $^{13}$C pocket, which is an uncertain parameter that is model and metallicity dependent \citep{Goriely:2004gw,Herwig:2004ki}.

\section{Stellar Modeling Uncertainties}\label{sec:modeluncertainties}
The uncertainties that have the greatest effect on the yields of heavy elements are the numerical treatments of convection, mass loss, and reaction rates, as well as the rotation in massive stars and low-temperature opacities in AGB stars \citep[e.g.,][]{Marigo:2002cu, Fishlock:2014hn,Constantino:2014gv}.

In stellar models of all masses, convective mixing plays a crucial role in the transport of energy and chemical species. The construction of accurate stellar models requires a method to approximate the effects of convection in 1D stellar evolution codes, as the high computational demands of full 3D hydrodynamical models limit their simulation times to no more than a small fraction of a stellar lifetime \citep[e.g.,][]{Stancliffe:2011dw}. The most common numerical treatment of convection is the mixing-length theory (MLT) that depends on the value of an uncertain parameter, $\alpha$, which is the mixing length in units of the local pressure scale height. The value of $\alpha$ is usually assumed to be constant on the AGB (e.g., the yields shown in Figure \ref{fig:agbyields} use a value of 1.86), even though empirical and theoretical studies both suggest that the value changes with stellar evolution \citep{Lebzelter:2007ho,Magic:2014uq}. Larger values of $\alpha$ have been shown to increase the depth of the third dredge-up \citep{Boothroyd:1988jl}, which increases the yields of $s$-process elements \citep{Cristallo:2009kn,Cristallo:2011fz}.

An alternative treatment of convective mixing that has been applied to AGB stars is the full-spectrum of turbulence \citep[FST;][]{Canuto:1991bj,Canuto:1996cq}. FST predicts a higher rate of energy transport than MLT, which leads to increased surface luminosities and higher interior temperatures in stellar models. In the intermediate-mass (4 to 6 M$_\sun$) models of \citet{Ventura:2008iv} that use FST, temperatures at the base of the convective envelope reach 90 to 110 MK, which is hot enough for extensive H-burning nucleosynthesis (hot bottom burning). Combined with a luminosity-dependent mass-loss law, the high luminosities of these models drive rapid mass-loss rates that shorten the thermally-pulsing AGB phase and reduce the number of dredge-up episodes \citep{Ventura:2005ge,Ventura:2013ct}. The limited dredge-up in these models leads to a negligible net yield of C+N+O in the stellar wind. Presumably, this would also result in negligible yields of $s$-process elements, although yields from an FST model with a full $s$-process network are, to our knowledge, not published at present.

Another major uncertainty in stellar modeling is the mass-loss rate and its dependence upon stellar parameters. Indeed, massive stars can lose more than half of their mass by the end of core He burning \citep{Chiosi:1986df}. In massive stars with rotation, mass loss transports angular momentum away from the stellar surface \citep[][to which we refer for details of the mass-loss prescription used in our massive star models]{Hirschi:2007br}. With the very low mass-loss rates expected at metallicities of $Z=10^{-5}$ ([Fe/H]$=-3.8$) and below, extremely metal-poor massive stars will evolve differently from observable OB stars \citep{Maeder:2000br}. Adding further complexity, the mass-loss rate would also be increased by the presence of a binary companion. For AGB stars, mass loss is very difficult to determine empirically without an accurate understanding of the dust composition and detailed models of the radiative transfer physics. Because the rate of mass loss controls the amount of time spent on the AGB and the number of thermal pulses, changes to the mass-loss rate have a significant effect on the predictions of stellar yields. In our models, we use \citet{Vassiliadis:1993jk} mass-loss rates along the AGB, which includes the switch to a superwind phase of extremely rapid mass loss near the tip of the AGB. An alternative is the \citet{Bloecker:1995ui} formula derived from dynamical calculations of the atmospheres of Mira-like stars, which predicts higher mass-loss rates and shorter AGB lifetimes.

Our massive star models include rotationally-induced mixing in the form of meridional circulation and shear instabilities which dramatically alter the yields of CNO and $s$-process elements, depending on the rate of rotation \citep{Frischknecht:2012il}. The best constraints on the rotation rates of low-metallicity massive stars come from the comparison of chemical signatures in low-metallicity, low-mass stars with the predictions of rotating stellar models. In order to explain the existence of high N/O and C/O ratios at times too early for AGB stars to contribute, \citet{Chiappini:2006km} infer rotation rates of around 0.5 times the break-up velocity ($v_{rot}/v_{crit} = 0.5$) at [Fe/H]$<-3$. \citet{Chiappini:2008fe} claim that rotation is independently supported by the low $^{12}$C/$^{13}$C ratios of metal-poor stars, which they report are consistent with models having rotational velocities of $v_{rot}/v_{crit} \simeq$ 0.5 to 0.6. \citet{Fabbian:2009jh} reach a less-certain conclusion about rotation and interpret high C/O ratios as possible signatures of either Population III stars or rotating Population II stars. The effect of rotation on the $s$-process yields is illustrated in Figure \ref{fig:rmsyields} which shows that rotation is the dominant effect.

For rotating massive star models, another uncertainty with an effect on $s$-process yield predictions is the competition between the $^{17}$O($\alpha$,$\gamma$)$^{21}$Ne and $^{17}$O($\alpha$,n)$^{20}$Ne reactions. This is because $^{16}$O is highly effective at capturing free neutrons, which produces $^{17}$O. Neutrons are then either recycled via $^{17}$O($\alpha$,n)$^{20}$Ne or lost via $^{17}$O($\alpha$,$\gamma$)$^{21}$Ne. The rate of the $^{17}$O($\alpha$,$\gamma$)$^{21}$Ne reaction is particularly uncertain at the relatively low energies of stellar interiors. The first experimental rates for this reaction were published by \citet{Caughlan:1988dn} (hereafter CF88) and subsequently disputed by \citet{Descouvemont:1993kc}, who predicted on theoretical grounds that the rate should be lowered by roughly a factor of 1000. However, more recent experimental work by \citet{Best:2011gb} supports a rate similar to CF88. \citet{Best:2013gb} report that the ratio between the ($\alpha$,$\gamma$) and ($\alpha$,n) reactions is best matched by using the CF88 rate divided by ten for $^{17}$O($\alpha$,$\gamma$)$^{21}$Ne and the \citet[][NACRE]{Angulo:1999kp} rate for $^{17}$O($\alpha$,n)$^{20}$Ne, the combination of which we will refer to as CF88/10 rates.

For $s$-process yields of both intermediate-mass AGB models and massive-star models, the $^{22}$Ne($\alpha$,n)$^{25}$Mg reaction plays a critical role in determining neutron fluxes, and for this reason it has been the subject of a number of studies \citep{Angulo:1999kp,Jaeger:2001ch,Koehler:2002cf,Karakas:2006fk}. Recent rates presented by \citet{Longland:2012ix} have reduced the uncertainties in AGB model abundances caused by uncertainty in these reactions to less than a factor of two.

A major uncertainty for the $s$-process in low-mass models concerns the formation of a $^{13}$C pocket. This is because the physical mechanism that leads to $^{13}$C pockets in stars is yet to be identified. Currently proposed candidates include convective-boundary mixing \citep{Herwig:2000ua,Cristallo:2004us}, rotational mixing \citep{Herwig:2001vb,Piersanti:2013dh}, or gravity-wave driven mixing \citep{Denissenkov:2003gx}. Eventually, a deeper understanding of the physics involved might completely eliminate the free parameter that determines the mass of the $^{13}$C pocket. At present, a variety of constraints have been derived from observations of carbon-enhanced metal poor stars \citep{Izzard:2009hg,Bisterzo:2012bf,Lugaro:2012ht}, planetary nebulae \citep{Shingles:2013kg,Miszalski:2013gi}, and post-AGB stars \citep{BonacicMarinovic:2007jl,DeSmedt:2012dp}.

Aside from the uncertain size of the partial mixing zone and resulting $^{13}$C pocket, an additional uncertainty relates to the stellar initial masses in which a $^{13}$C pocket can be formed. With increasing stellar mass, the size of the He-rich intershell region decreases and temperatures at the base of the convective envelope during the third dredge-up increase, inhibiting $^{13}$C-pocket formation in more massive AGB stars. \citet{Goriely:2004gw} show that when the third dredge-up takes place with temperatures of around 40 to 70 MK $^{13}$C-pocket formation can be suppressed, depending on the details of any diffusive mixing near the convective boundary. At our metallicity of $Z=0.001$, the results of \citet{Goriely:2004gw} suggest that $^{13}$C-pocket formation could become inhibited above around 3.0 to 3.5 M$_\sun$ (but see \citet{Straniero:2014jk} for a different view on $^{13}$C-pocket formation above this mass). To account for this uncertainty on our results, we separately consider two cases in which our nucleosynthesis post process includes a PMZ for all stellar masses up to 3.0 or 3.5 M$_\sun$. This is an approximation in the absence of a physically-accurate PMZ included in our stellar model calculations.

In this work we do not consider binary stars, although the presence of a binary companion will also alter the yields with a dependence on the period and mass ratio of the system. 

\section{Observational Data}
\subsection{Differential abundances and empirical $s$-process distributions}
As an indication of how elemental abundances vary between two stars or stellar populations, it is common to subtract solar bracket [X/Fe] abundances \citep[e.g.,][]{Yong:2008in,Roederer:2011hw}. The difference [X/Fe]$_2 - $[X/Fe]$_1$ is equal to $\log_{10} [($X$/$Fe$)_2/($X$/$Fe$)_1]$, i.e., it measures of the number ratio of X to Fe in system 2 as a factor of the ratio in system 1. In the case that system 1 represents an initial composition that has undergone nucleosynthesis to make the abundances in system 2, a quantity that isolates the net production or destruction of elements is obtained by subtracting the number ratios in linear abundance space, i.e., $\Delta$(X/Fe)=$($X/Fe$)_2 - ($X/Fe$)_1$, assuming that Fe is either constant or only marginally produced. This quantity is analogous to the net yields of stellar nucleosynthesis models, which are computed by subtracting the abundances in the initial composition from the abundances in the stellar ejecta \citep[e.g.,][]{Karakas:2010et}.

Using a linear abundance subtraction, \citet[][Table 8]{Roederer:2011hw} calculate an $s$-process-only residual composition for M22 by subtracting the average X/H number ratios of $s$-poor from $s$-rich stars. We use the same technique to derive empirical $s$-process distributions for M4 and M22, except that we use number ratios relative to Fe. Our own testing confirms that the resulting distributions look very similar regardless of whether abundances relative to Fe or H are used.

\begin{figure}
 \begin{center}\includegraphics[width=1.0\columnwidth]{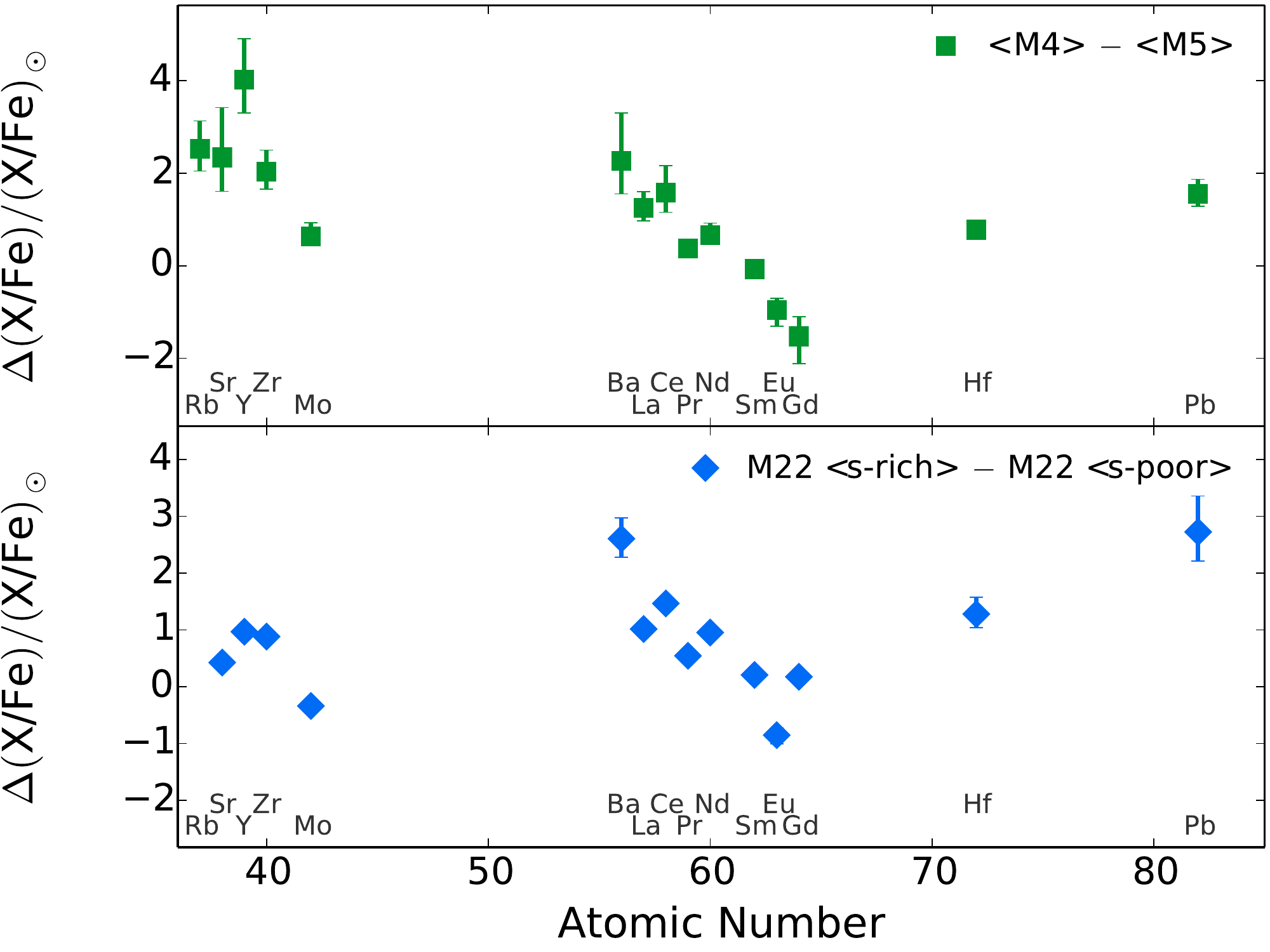}\end{center}
 \caption{Abundance differences relative to solar with observational data. Abundances of M4 and M5 are from \citet{Yong:2008in,Yong:2008ib} except Cu from \citet{Simmerer:2003kg} and Ba from \citet{Ivans:2001ju}. M22 abundances are from \citet{Roederer:2011hw}. $\Delta$(X/Fe)=$($X/Fe$)_A - ($X/Fe$)_B$. Upper and lower bounds are calculated by multiplying and dividing by $10^{\sqrt{(\sigma A)^2+(\sigma B)^2}}$, where $\sigma A$ and $\sigma B$ are the logarithmic abundance dispersions of two systems whose abundances have been subtracted.}
 \label{fig:observations}
\end{figure}

Figure \ref{fig:observations} shows our calculated $s$-process-only residuals of M22 $<$s-rich$>$ $-$ $<$s-poor$>$ and $<$M4$>$ $-$ $<$M5$>$ relative to the solar abundances \citep{Asplund:2009eu}. To visually emphasize small differences (that are significant within the errors), we plot on a linear scale. The distance from the zero point is related to the amount of dilution with $s$-poor material, while the shape of the distribution is relatively independent of this uncertain parameter and primarily depends on the relative abundances in the stellar ejecta. In agreement with \citet{Roederer:2011hw}, we interpret the empirical $s$-process distributions of M4 and M22 as representing enrichment by material of a similar but not identical composition. We suggest that the $s$-process distributions of these two clusters are distinguishable as representing the results of different nucleosynthetic sites or stellar mass ranges. We now discuss the observations, starting with M4.

\subsection{The s-rich globular cluster M4}
M4 is a typical mono-metallic globular cluster with a Na-O anti-correlation and constant abundances of Fe-group elements, neutron-capture elements \citep[except possibly Y, see][]{Villanova:2011fo}, and C+N+O \citep{Drake:1992kw,DOrazi:2010kb,Marino:2008du,Marino:2011hu}.

Although the neutron-capture element abundances show no star-to-star variations in M4, the entire cluster is moderately enriched with $s$-process elements compared to other globular clusters at a similar metallicity, such as M5. With [Fe/H] of $-1.33$ \citep{Carretta:2009di}, M5 is a near metallicity-twin of M4 with similar abundances of Fe-peak (Fe, Co, Ni) and $r$-process (Eu) elements. Compared to M5, the $s$-process elements in M4 are overabundant by between 0.3 and 0.5 dex \citep{Ivans:2001ju,Yong:2008in,Yong:2008ib}. Figure \ref{fig:observations} shows that the $s$-process distribution of M4 (which is obtained by subtracting the abundances of M5) is dominated by the light $s$-peak around Y, with lower abundances of Ba and heavier $s$-process elements.

\citet{Karakas:2010wy} and \citet{Roederer:2011hw} have suggested that the overabundances of Rb, Y, Ba, La, and Pb in M4 relative to M5 could result from intermediate-mass AGB stars (in which the neutron source is $^{22}$Ne($\alpha$,n)$^{25}$Mg) by a comparison with individual stellar yields. However, this does not rule out a simultaneous contribution from less massive stars with $^{13}$C pockets. The simultaneous contribution of the $s$-process from both $^{13}$C pockets and the $^{22}$Ne source is the conclusion drawn by \citet{Straniero:2014jk}, who fit to the $s$-process distribution of M4 to an IMF-weighted sum of stellar yields with AGB models from 3 to 6 M$_\odot$ at [Fe/H]$=-1.8$.

Although AGB stars have been suggested as the heavy elements producers in M4, the sequence of events that led to the peculiar $s$-process enrichment of M4 and not M5 (and many other GCs) is presently without a conclusive explanation in the literature.

\subsection{The two populations in M22}

\citet{Marino:2009je} demonstrated that M22 exhibits two groups of stars separated by 0.15 dex in [Fe/H] and variations in $s$-process elements that are correlated with Fe. 

The first group ($s$-poor) has a mean metallicity of [Fe/H]$=-1.82 \pm 0.02$ and [$s$/Fe] of $-0.01 \pm 0.01$, where $s$ represents an average over Y, Zr, Ba, La, and Nd. The second group ($s$-rich) has a metallicity of [Fe/H]$=-1.67 \pm 0.01$ and [$s$/Fe] of $+0.35 \pm 0.02$ \citep{Marino:2011kz}. Both populations independently show the Na-O and C-N anti-correlations \citep{Marino:2011kz}, indicating that whichever stars contributed to the enrichment of the $s$-rich population did not also produce the light element anomalies. In comparison with M4, which is mostly enriched with light $s$-elements, Figure \ref{fig:observations} shows that the $s$-process distribution of M22 is peaked at the heavy $s$-elements near Ba.

\citet{Marino:2012db} compare photometry of the two groups with isochrones and derive an upper-limit age spread of 300 Myr. The result is confirmed by \citet{Joo:2013dr}, who find that their best-fitting isochrones predict an age difference of $0.3 \pm 0.4$ Gyr. Assuming that the gas cooling time is a negligible fraction of a stellar lifetime, the age difference of 300 Myr allows enough time for stellar masses as low as 3.0 M$_\sun$ to contribute to the chemical abundances in the $s$-rich group. The connection between the minimum contributing mass and the timescale for $s$-process enrichment is explored in more detail in Section \ref{sec:discussion}.

\section{Chemical Evolution Model and Results}\label{sec:cemresults}
We present abundance evolution results calculated using a new code, Evel ChemEvol to solve the equations of chemical evolution for a single-zone \citep[for an review, we refer to][]{Pagel:2009ws}. Our testing with the AGB yields and self-pollution scenario described by \citet{Fenner:2004ju} confirms that the code correctly reproduces the abundance results of an existing chemical evolution code. For more details of the output validation tests, see Appendix \ref{appendix:code}.

Our simplified chemical evolution model includes a single short burst of star formation as a first-order attempt at understanding the enrichment of globular clusters. The final abundance outputs of the chemical evolution model represent the IMF-weighted \citep{Kroupa:1993tm} sum of ejecta from a range of stellar masses with yields that are interpolated from a grid of stellar models.

Our derivation of an $s$-process-only component from the observational abundances and the similar subtraction of the initial abundances from the final abundances of the models (or the subtraction of the final abundances of two different models) enables us to compare our chemical evolution predictions with both cluster systems simultaneously, although the initial composition will affect the ratios of elements in the stellar yields (e.g., [ls/hs]).

For each stellar mass in the range from 15 to 40 M$_\odot$, massive star yields are interpolated from our grid of stellar models with initial rotation rates of 0.0 and 0.4 as a fraction of $v_{crit}$ \citep{Frischknecht:2012il}. For the particular initial mass of 25 M$_\sun$, we have also have yields from stellar models with rotation rates of 0.4 and 0.5 $v_{crit}$ with and without alternative reaction rates (CF88/10). From the yields of the 25 M$_\sun$ stellar models, we calculate a set of factors (one per chemical species) that approximate the effect of these alternative parameters on the yields of the other models in the grid with different initial masses. 

Table \ref{tab:allresults} and Figures \ref{fig:rmszm3results}, \ref{fig:rmszm5results}, and \ref{fig:agbresults} show the quantitative results of our chemical evolution models with rotating massive stars and AGB stars. The first two rows show the observation results in terms of [ls/hs]\footnote{We define [ls/Fe] = ([Y/Fe] + [Zr/Fe])/2, [hs/Fe] = ([Ba/Fe] + [La/Fe] + [Ce/Fe])/3, and [ls/hs] = [ls/Fe] - [hs/Fe].} and [Pb/hs] ratios of the $s$-process residuals. A timescale is given for the models with AGB stars, which is the stellar lifetime of the lowest included mass.

\begin{table*}
\begin{center}\caption{Observational $s$-process residuals and the results of our chemical evolution models as well as individual AGB yields. The CF88/10 case is explained in the Section \ref{sec:cemresults}.\label{tab:allresults}}
  \vspace{1mm}
\begin{tabular}{l l r r r}
\tableline\tableline
A						& B				& [ls/hs]$_{A-B}$	& [Pb/hs]$_{A-B}$	& Timescale (Myr)\\
\hline
  M4 						& M5 			&  0.24			& $-0.03$	& -\\
  M22 ($s$-rich) 			& M22 ($s$-poor)	& $-0.23$			& 0.24	& $\sim$ 300\footnote{Derived from isochrone fitting of the subgiant branch region by \citet{Marino:2012db}.}\\
\tableline
\multicolumn{5}{l}{Results with [Fe/H] $=-3.8$ rotating massive star yields from 15 to 40 M$_\odot$}\\
\tableline
RMS\footnote{Rotating massive stars.}  ($v_{rot}=0.5$, CF88/10) & RMS($v_{rot}=0.4$, CF88/10)	& 0.82	& $-2.31$	& -\\
RMS ($v_{rot}=0.5$, CF88/10) & RMS ($v_{rot}=0.0$)			& 1.26	& $-2.02$	& -\\
RMS ($v_{rot}=0.5$) 		& RMS ($v_{rot}=0.4$)					& 2.73	& $-0.77$	& -\\
RMS ($v_{rot}=0.5$) 		& RMS ($v_{rot}=0.0$)					& 2.62	& $-0.50$	& -\\
\tableline
\multicolumn{4}{l}{Results with [Fe/H] $=-1.8$ rotating massive star yields from 15 to 40 M$_\odot$}\\
\tableline
RMS ($v_{rot}=0.4v_{crit}$)	 & [Fe/H] $=-1.8$ ($\alpha$-enhanced scaled solar)	& 1.95	& $-0.11$	& 12\\
\tableline
\multicolumn{5}{l}{Results with [Fe/H] $=-1.2$ AGB yields (M $\leq 3.0$ M$_\odot$ stellar models include a PMZ)}\\
\tableline
AGB 3.50 to 7.0 M$_\odot$	& [Fe/H] $=-1.2$ (scaled solar) 		& 0.72	& $-0.62$	& 199\\
AGB 3.25 to 7.0 M$_\odot$	& [Fe/H] $=-1.2$ (scaled solar)		& 0.73	& $-0.61$	& 239\\
AGB 3.00 to 7.0 M$_\odot$	& [Fe/H] $=-1.2$ (scaled solar)		& $-0.01$	& 0.72	& 290\\
AGB 2.75 to 7.0 M$_\odot$	& [Fe/H] $=-1.2$ (scaled solar)		& $-0.30$	& 0.79	& 364\\
\tableline
\multicolumn{5}{l}{Results with [Fe/H] $=-1.2$ AGB yields (M $\leq 3.5$ M$_\odot$ stellar models include a PMZ)}\\
\tableline
AGB 4.00 to 7.0 M$_\odot$	& [Fe/H] $=-1.2$ (scaled solar)		& 0.72	& $-0.62$	& 144\\
AGB 3.50 to 7.0 M$_\odot$	& [Fe/H] $=-1.2$ (scaled solar)		& 0.09	& 0.86	& 199\\
AGB 3.25 to 7.0 M$_\odot$	& [Fe/H] $=-1.2$ (scaled solar)		& $-0.10$	& 0.92	& 239\\
AGB 3.00 to 7.0 M$_\odot$	& [Fe/H] $=-1.2$ (scaled solar)		& $-0.25$	& 0.89	& 290\\
AGB 2.75 to 7.0 M$_\odot$	& [Fe/H] $=-1.2$ (scaled solar)		& $-0.37$	& 0.86	& 364\\
\tableline
\multicolumn{5}{l}{[Fe/H] $=-1.2$ individual-mass AGB yields}\\
\tableline
AGB 3.50 M$_\odot$			& [Fe/H] $=-1.2$ (scaled solar)	& 0.94	& $-0.34$	& 199\\
AGB 3.25 M$_\odot$			& [Fe/H] $=-1.2$ (scaled solar)	& 0.94	& $-0.30$	& 239\\
AGB 3.00 M$_\odot$ w/ PMZ		& [Fe/H] $=-1.2$ (scaled solar)	& $-0.56$	& 0.78	& 290\\
AGB 2.75 M$_\odot$ w/ PMZ		& [Fe/H] $=-1.2$ (scaled solar)	& $-0.55$	& 0.84	& 364\\
\tableline
\end{tabular}\end{center}
\end{table*}

\subsection{Rotating Massive Stars}
\begin{figure}
 \begin{center}\includegraphics[width=1.0\columnwidth]{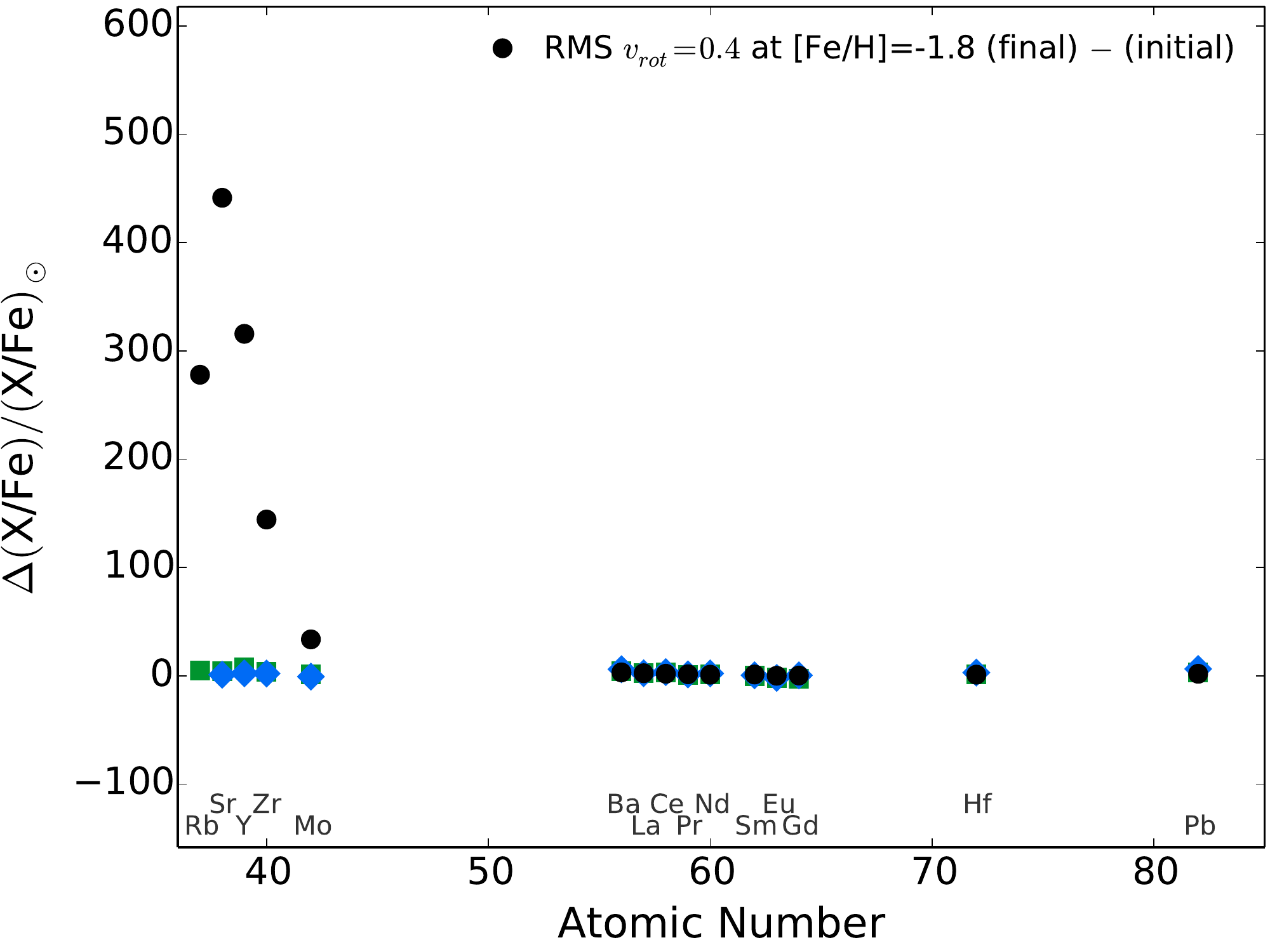}\end{center}
 \caption{Chemical evolution results for rotating massive star yields at [Fe/H]$=-1.8$ with rotation rates at 40\% of the break-up velocity. Also shown are the empirical distributions of M4 (green) and M22 (blue) scaled to match La abundance.}
 \label{fig:rmszm3results}
\end{figure}

\begin{figure}
 \begin{center}\includegraphics[width=1.0\columnwidth]{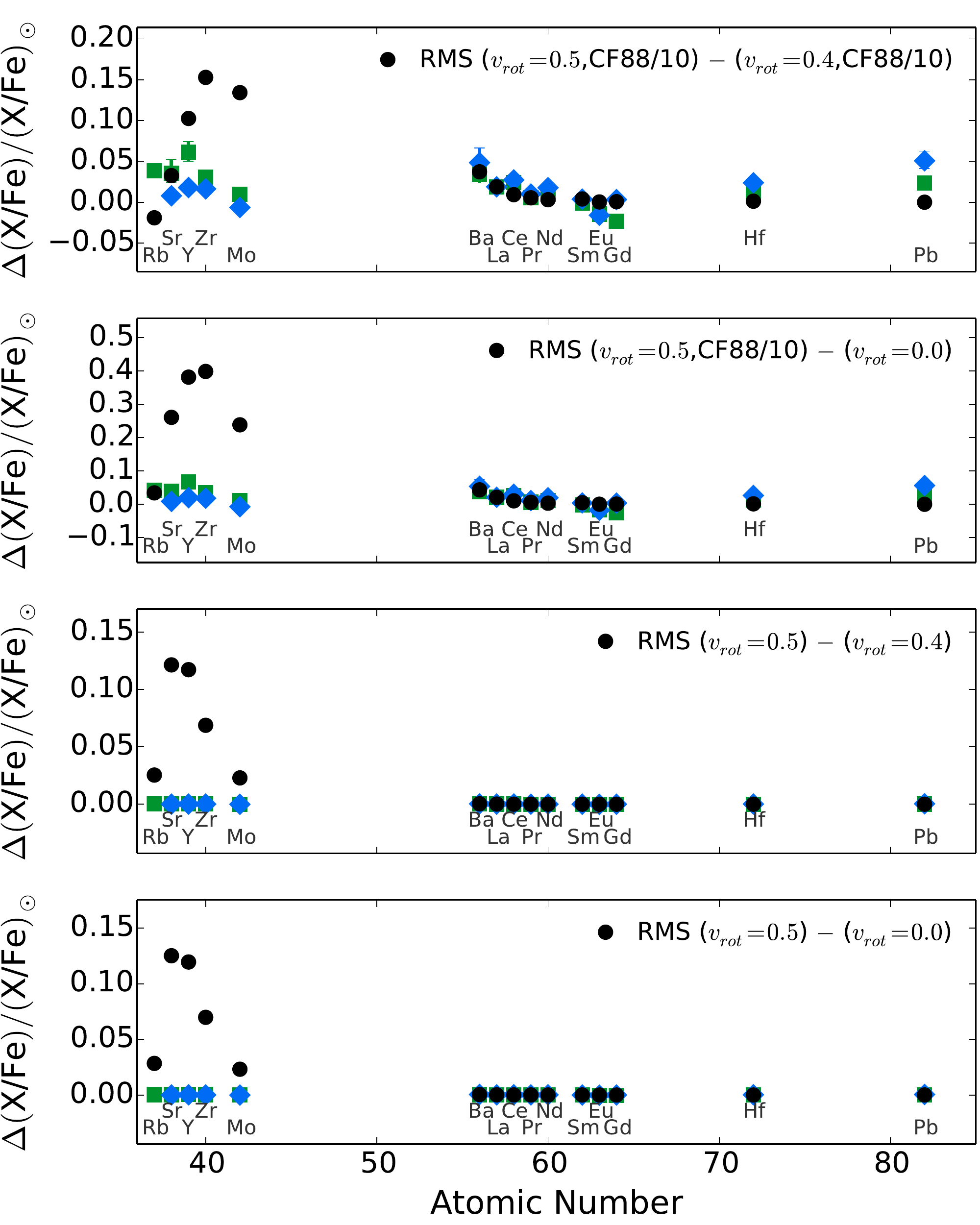}\end{center}
 \caption{Chemical evolution abundance subtraction results for rotating massive star models at [Fe/H]$=-3.8$ with rotation rates of 0\%, 40\%, and 50\% of the break-up velocity and an alternative reaction rate (CF88/10). Also shown are the empirical distributions of M4 (green) and M22 (blue) scaled to match La abundance.}
 \label{fig:rmszm5results}
\end{figure}

Figure \ref{fig:rmszm5results} shows the chemical evolution results for rotating massive stars at very low metallicity ($Z=10^{-5}$), where the abundances of models with low or no rotation have been subtracted from the abundances of faster rotating models to derive an $s$-process residual. These results correspond to the scenario of stochastic enrichment in which early generations of massive stars that formed M4 and M5 had a higher average rotation rate in the case of M4. For M22, these results correspond to a scenario in which the two groups chemically evolved separately. Although we only consider yields with a single value of $v_{rot}$ in each chemical evolution model, future studies that model a distribution of rotational velocities would be of great interest. The resulting distributions are a poor match to the empirical distributions of both M4 and M22 (Figure \ref{fig:observations}), as they predict a very strong weighting towards light $s$-peak elements, even using the highest rotation rates and with an alternative reaction rate (CF88/10) that limits the effectiveness of $^{16}$O as a neutron poison. The poor match to observations is also apparent from the high [ls/hs] ratios of 0.8 to 2.7 shown in Table \ref{tab:allresults}, as compared with 0.24 in M4 $-$ M5 and $-0.23$ in M22 $s$-rich $-$ $s$-poor.

To test the scenario for M22 in which rotating massive stars of the $s$-poor group have driven the increase in both [Fe/H] and the $s$-process abundances in the $s$-rich group, we present chemical evolution results from a generation of rotating massive stars at [Fe/H]$=-1.8$ that are shown in Figure \ref{fig:rmszm3results}. The abundances of the initial composition have been subtracted from the final (ejecta) abundances to derive an $s$-process residual using the same technique applied to M4 and M22. The $s$-process distribution is too strongly weighted toward elements at the first peak around Y (with an [ls/hs] ratio of 1.95) to match the observational distribution of M22.

\subsection{AGB Stars}
\begin{figure}
 \begin{center}\includegraphics[width=1.0\columnwidth]{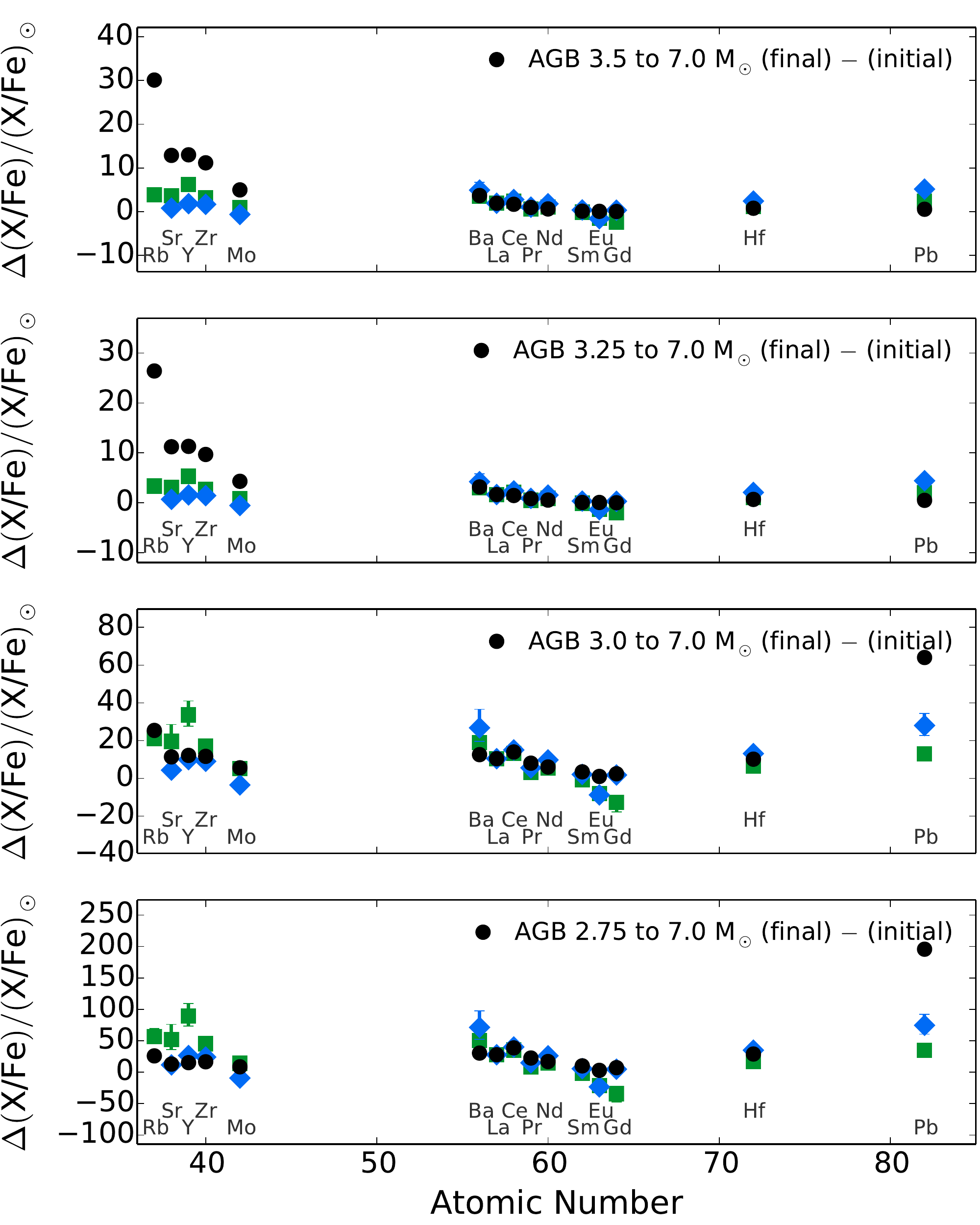}\end{center}
 \caption{Chemical evolution results with several mass ranges of AGB yields at [Fe/H]$=-1.2$, and where the highest mass to include partial mixing zone is 3.0 M$_\odot$. Also shown are the empirical distributions of M4 (green) and M22 (blue) scaled to match La abundance.}
 \label{fig:agbresults}
\end{figure}

\begin{figure}
 \begin{center}\includegraphics[width=1.0\columnwidth]{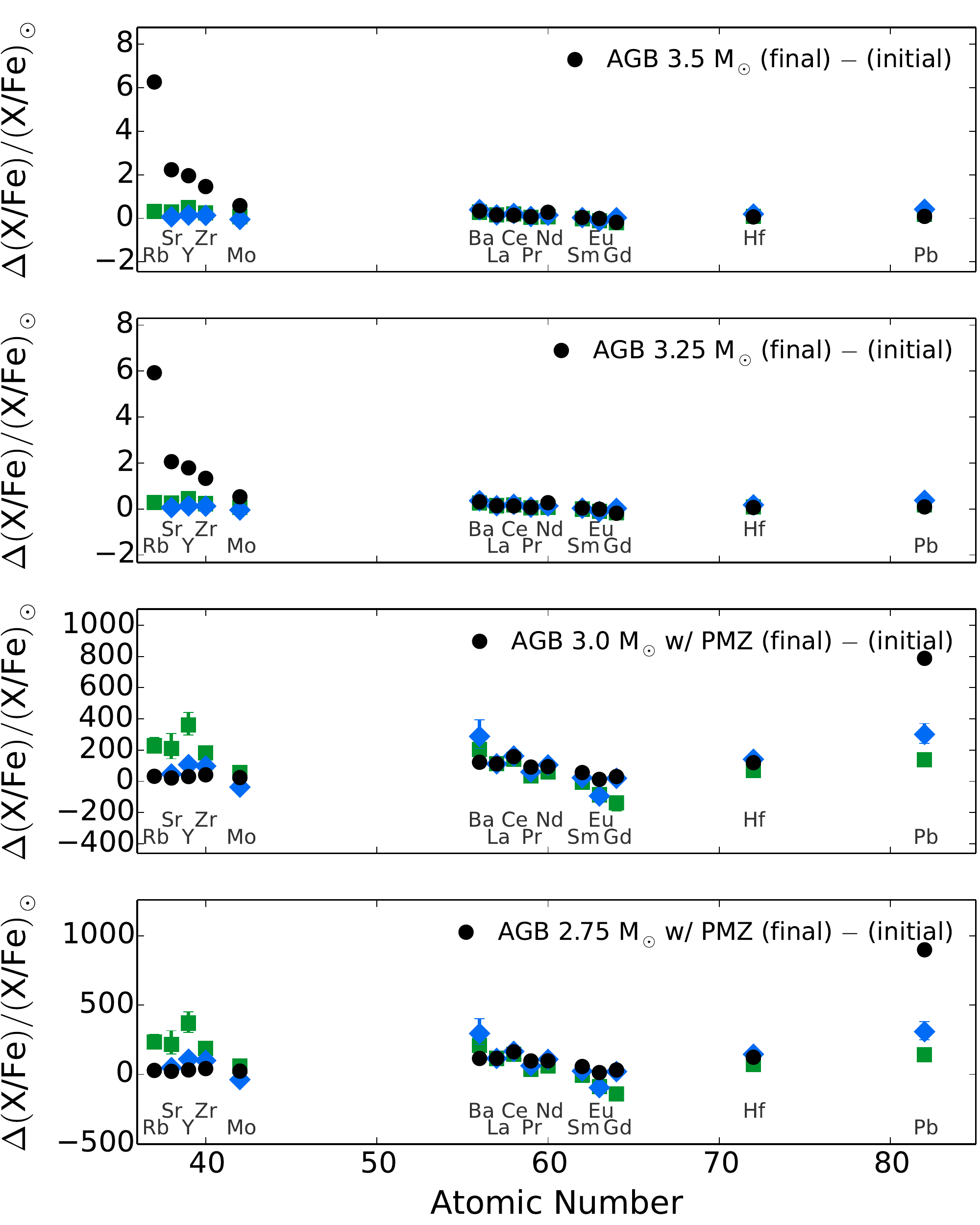}\end{center}
 \caption{Results with single-mass AGB yields at [Fe/H]$=-1.2$. Also shown are the empirical distributions of M4 (green) and M22 (blue) scaled to match La abundance.}
 \label{fig:agbindividualresults}
\end{figure}

We test chemical evolution models that predict the output of a single generation of low-metallicity AGB stars, with the results provided in Figure \ref{fig:agbresults} and Table \ref{tab:allresults}. We vary the lower limit of the stellar mass range as a free parameter because this corresponds to the uncertain age difference between the $s$-process polluters and the $s$-process-rich stars (minus the gas cooling time). Because of the uncertainty over the upper mass limit for AGB stars to have a $^{13}$C pocket, we separately test chemical evolution models in which the 3.25 and 3.5 M$_\odot$ yields are calculated from models with and without a PMZ of $1 \times 10^{-3}$ M$_\odot$.

For M4, the [ls/hs] and [Pb/hs] ratios are bracketed from above and below by models with AGB yields that have lower limit masses of 3.00 and 3.25 M$_\odot$, respectively. From the stellar lifetimes, this corresponds to a minimum enrichment timescale 239-290 Myr. As the 3.00 M$_\odot$ stellar model includes a PMZ and the 3.25 M$_\odot$ model does not, this indicates a small contribution from stars with a $^{13}$C pocket. If the models up to 3.5 M$_\odot$ include a PMZ, the [ls/hs] and [Pb/hs] ratios of M4 are bracketed by 3.5 and 4.0 M$_\odot$ lower-limit models, corresponding to a 140-200 Myr minimum enrichment timescale. With the uncertain upper mass limit for the $^{13}$C pocket formation, the minimum enrichment timescale for M4 is likely around 140-290 Myr.

Although our AGB yields are not an exact match to the metallicity of M22 ([Fe/H] $=-1.2$ versus $-1.8$ in M22's $s$-poor group), we explore the similarities between our chemical evolution results and the observational data. The chemical evolution of heavy elements in the $s$-rich group of M22 requires even lower mass stars than M4, however a simultaneous match of [ls/hs] and [Pb/hs] ratios is not found in our results. The [ls/hs] ratio in M22 is bracketed by 2.75 and 3.0 M$_\odot$ lower-limit models, while the [Pb/hs] ratio is bracketed by models with lower mass limits of 3.00 and 3.25 M$_\odot$. If the stellar masses up to 3.5 M$_\odot$ include a PMZ, we find that M22's [ls/hs] ratio is between those of the 3.00 and 3.25 M$_\odot$ lower-limit models, while the predicted [Pb/hs] of these models is still too high to match the data. In both of our test cases for the upper limit mass of $^{13}$C pocket formation, the dual contribution from stars with a $^{13}$C pocket as well as stars with a $^{22}$Ne neutron source are required. With our assumption that $^{13}$C pockets transition from fully developed to negligible between initial masses of 3.0 and 3.5 M$_\sun$, we predict a lower limit on the polluter masses of 2.75 to 3.25 M$_\odot$, which corresponds to a minimum enrichment timescale of 240-360 Myr.

A common method for comparing measured abundances with the predictions of stellar models is to use the yield results of a single stellar model rather than a grid covering a range of stellar masses that has been weighted by an initial mass function. In Table \ref{tab:allresults} and Figure \ref{fig:agbindividualresults}, we present single-mass yield results for comparison with our chemical evolution results. The slope of the IMF means that the lowest contributing mass will have the largest contribution to the final abundances, however the single models with a $^{13}$C pocket importantly lack the significant production of light $s$-elements that is due to intermediate-mass AGB stars. This difference is apparent in the high Rb/Sr ratio and overall higher abundances of Rb, Sr, Y, Zr, and Mo by the 3.0 to 7.0 M$_\sun$ model shown in Figure \ref{fig:agbresults}, as compared with same ratio from the single 3.0 M$_\sun$ model shown in Figure \ref{fig:agbindividualresults}. 

\section{Discussion and Conclusions}\label{sec:discussion}
We have used the shape of the $s$-process distributions in M4 and M22 to identify the $s$-process polluter mass range and the corresponding maximum stellar lifetime, which places a lower limit on the timescale of $s$-process enrichment. As well as $s$-process enrichment, M4 and the two groups in M22 also feature anti-correlated variations in O and Na, which we do not attempt to explain. Although we match the $s$-process distribution of M22's $s$-rich group with the ejecta of AGB stars, the coexistence of an Fe variation in M22 likely required some fraction of the ejecta from massive stars to be kept within the cluster to form new stars with a higher Fe abundance.

The lower-limit mass range of 2.75-3.25 M$_\odot$ in our best-fitting models for M22 corresponds to a stellar lifetime in the range $300 \pm 60$ Myr. Assuming that the time for the ejecta to cool and form new stars is relatively small, this value is consistent with the 300 Myr upper limit derived from isochrone fitting of the subgiant branch region by \citet{Marino:2012db} and \citet{Joo:2013dr}. A match between the inter-group age difference and the lifetime of the minimum polluter mass supports a scenario in which the metal-rich group in M22 has been self-enriched with material ejected from stars coeval with the present-day $s$-poor group. Alternative scenarios in which the two metallicity groups in M22 (which have independent light element anti-correlations) are the result of a merger of two separate GC systems or the second generation is formed from $s$-process rich material accreted from outside the cluster are also plausible. However, under both of these alternative scenarios, the close match between timescales of pollution and the age difference between the stellar groups would be a coincidence.

Further evidence for a lower mass limit of $\approx 3$ M$_\sun$ and an enrichment timescale of $\approx300$ Myr for M22 is the measured 0.6 dex spread in F abundances \citep{AlvesBrito:2012hy,DOrazi:2013dm}. \citet{DOrazi:2013dm} report F abundances that correlate with O, are anti-correlated with Na, and increase between the two groups. The authors suggest that the $s$-rich group has been enriched by the ejecta of stars with masses between 4 and 5 M$_\sun$, as these stars would destroy (rather than produce) F while O is destroyed in the early stages of GC formation. However, these measurements could be heavily affected by systematic errors as is claimed by \citet{deLaverny:2013it}, who argue that a reliable detection of the HF line in M22 stars is unlikely due to errors in radial velocity correction, continuum subtraction, and the removal of telluric absorption lines.

The matching of M4 $-$ M5 to a model of AGB ejecta opens up the question of how the formation of M4 differed to that of M5. A scenario similar to M22 in which an $s$-poor generation of stars pollute the interstellar medium from which a second generation forms is ruled out by observations of constant $s$-process abundances in M4, which do not feature the same bimodality found in M22. A more likely scenario is that M4 and M5 formed out of material in an inhomogeneous early Galactic halo. \citet{James:2004br} show that Ba and Eu abundances plotted as a function of [Fe/H] for mono-metallic globular clusters (including M4 and M5) fall within the spread of halo field star values, suggesting that they share a common origin or a similar enrichment process.

Our inferred enrichment timescales for M4 and M22 are roughly a factor of two larger than the $150 \pm 50$ Myr for both clusters inferred by \citet{Straniero:2014jk}. They require a larger minimum contributing mass ($4.0 \pm 0.5$ M$_\odot$) due to their inclusion of a prescription for core-envelope convective boundary mixing \citep{Cristallo:2009kn} that predicts small $^{13}$C pockets in AGB models with masses as high as 4.5 M$_\odot$. The predictions of \citet{Straniero:2014jk} and those in this paper are both consistent within the uncertainty of the age spread in M22 derived from isochrone fitting. Our results support their conclusion that neutron captures from both $^{13}$C pockets and the $^{22}$Ne source operating in convective pulses are required to explain the $s$-process enrichment of M4 and M22.

While our massive star models could not reproduce the $s$-process enhancements seen in M4 and in M22, there are still large uncertainties on the yields of $s$-process elements from rotating massive star models. For example, the yields of \citet{Pignatari:2008ec} show a ratio between Y and Ba of approximately unity (see their Figure 2). This suggests that while AGB stars produce the best fit with our adopted stellar yields, other sets of yields may change our conclusions as to the nature of the polluters of heavy elements in globular clusters.

We consider the effect of a possible $r$-process difference between the $s$-rich and $s$-poor samples of up to [r/Fe] $=0.4$. Using the solar system $r$-process fractions of \citet{Simmerer:2004ib}, the effect would be to increase [Y/Fe] by 0.15 dex, [Zr/Fe] by 0.07 dex, [Ba/Fe] by 0.09 dex, [La/Fe] by 0.14 dex, and [Ce/Fe] by 0.11 dex. The net result for the [ls/hs] ratio would be a change of less than 0.01 dex. The ratio [Pb/Fe] would increase by 0.12 dex, resulting in a [Pb/hs] change of less than 0.01 dex. We conclude that our results hold independently of a possible $r$-process difference between M4 and M5 or the two populations in M22. A dilution by pristine material would shift the [X/Fe] ratios in the $s$-process residual to closer to zero, but would not affect the relative abundances between elements.

Our models predict [Pb/hs] ratios that are too high to match the observations of M22. A similar phenomenon is reported by \citet{DeSmedt:2014hd} for metal-poor ([Fe/H]$<-1$) post-AGB stars in the Magellanic clouds, which they refer to as the `lead discrepancy'. If the Pb measurements are correct, then a solution to the lead discrepancy will likely require a better understanding of the mixing that leads to the formation of a $^{13}$C pocket, possibly by modeling it as an advective process, rather than more typical diffusive treatment. One form of extra mixing that is not included in our AGB models (or in most AGB stellar models) is the mixing due to rotation. The study of AGB models with rotation by \citet{Piersanti:2013dh} hints at a possible solution to the lead discrepancy, as they find that rotation reduces the [Pb/hs] ratio in the stellar yields.

Rb is overproduced in our best-fitting model in comparison with the observational data for M4, while separate Rb abundances for the two groups in M22 are not available in the literature. An overproduction of Rb in AGB stellar models is also noted by \citet{DOrazi:2013jr} under the assumption that AGB stars are responsible for the light element variations in M4. Their 6 M$_\odot$ model with the mass-loss rate from \citet{Vassiliadis:1993jk} and a mixing-length parameter of $\alpha = 1.75$ produces too much variation in neutron-capture elements, for too little variation in Na. Their solution is to use the higher mass-loss rates of \citet{Bloecker:1995ui} and a boosted mixing-length parameter ($\alpha = 2.2$), which improves the fit to the abundances in M4 by increasing the temperature at the base of the convective envelope and reducing the cumulative dredge-up of $s$-process elements into the envelope.

The opposite case of a Rb underproduction is found when stellar models (5-9 M$_\odot$) are compared with AGB stars in the Galaxy and Magellanic clouds \citep{vanRaai:2012fq,Karakas:2012kc}, although recent work by \citet{Zamora:2014ch} suggests that the inferred Rb abundances may be systematically overestimated due to the presence of circumstellar envelopes. The implementation of a delayed superwind to increase Rb yields explored by \citet{Karakas:2012kc} would likely worsen the discrepancy in our results, unless there was a simultaneous reduction in Rb production by the less massive ($<5$ M$_\odot$) AGB models.

Future stellar models at the correct metallicity for M22, and more generally improvements to the numerical treatment of mixing and mass loss might help to reduce some of the discrepancies with observations of the $s$-process abundances in globular clusters. Even with present models, the application of similar techniques to other clusters with $s$-process variation such as M2 \citep{Yong:2014wd} and $\omega$ Centauri would enable us to characterize the full range of enrichment timescales and polluter masses among the anomalous globular clusters.

\acknowledgements{This research has made use of NASA's Astrophysics Data System. LJS and AIK thank Chris Sneden for helpful discussions about spectroscopic uncertainties. A.I.K. was supported through an Australian Research Council Future Fellowship (FT110100475). R.H. acknowledge the support from Eurocore project Eurogenesis and ERC Starting Grant No.~306901. R.H. and AIK acknowledge support from the World Premier International Research Center Initiative (WPI Initiative), MEXT, Japan.}

\bibliographystyle{apj}
\bibliography{references}

\appendix
\section{Verification of Evel ChemEvol code}\label{appendix:code}
To validate the output of the new chemical evolution code Evel ChemEvol used in this study, we use the stellar yields and globular cluster self-pollution scenario described by \citet{Fenner:2004ju}. A metal-free initial composition is first polluted with the ejecta of massive stars up to a metallicity of [Fe/H]$=-1.4$. Subsequently, star formation takes place on a timescale of $10^7$ years. The ejecta from stars $< 6$ M$_\sun$ is kept within the system, while ejecta from more massive stars is lost.

\begin{figure}
 \begin{center}\includegraphics[width=0.45\columnwidth]{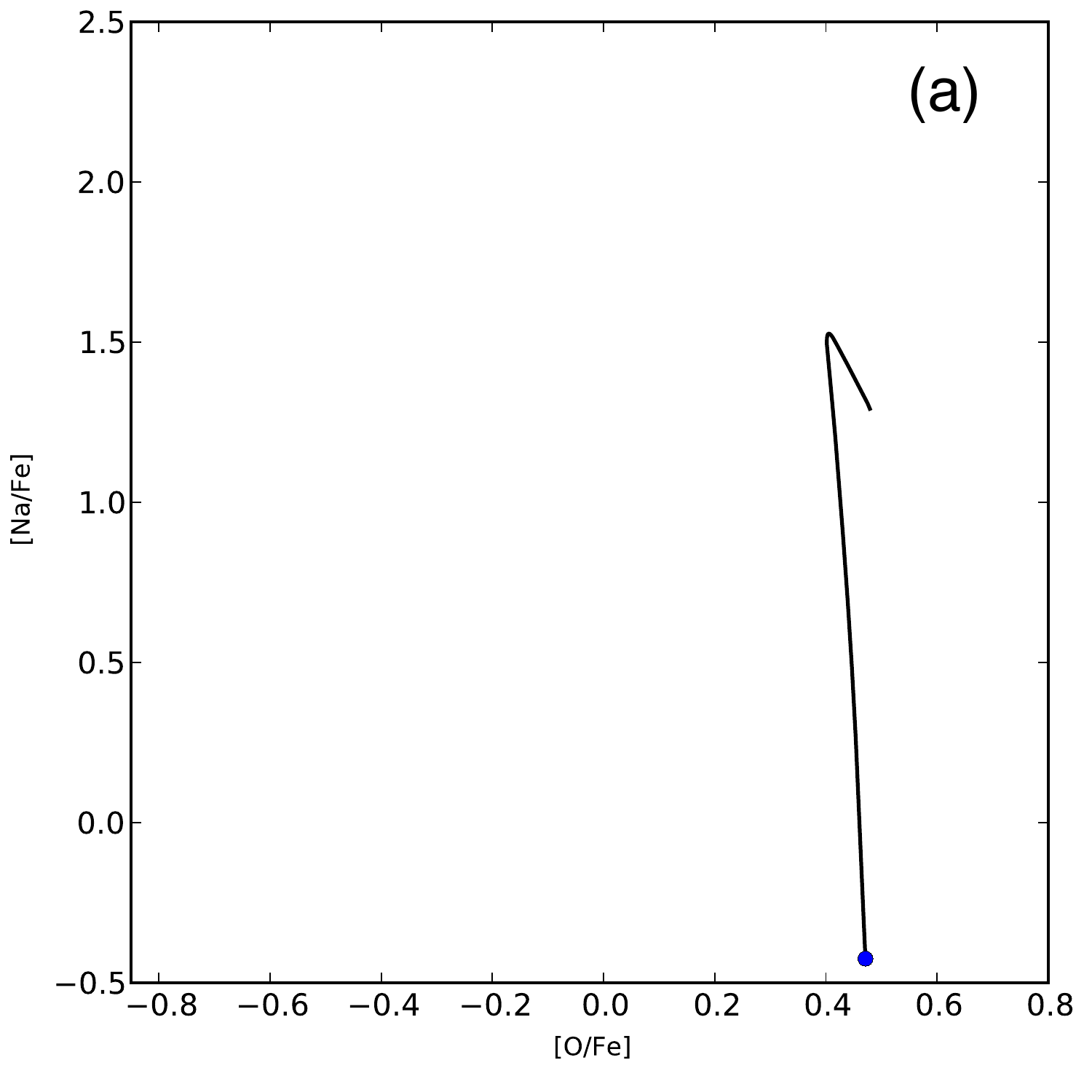}  \includegraphics[width=0.45\columnwidth]{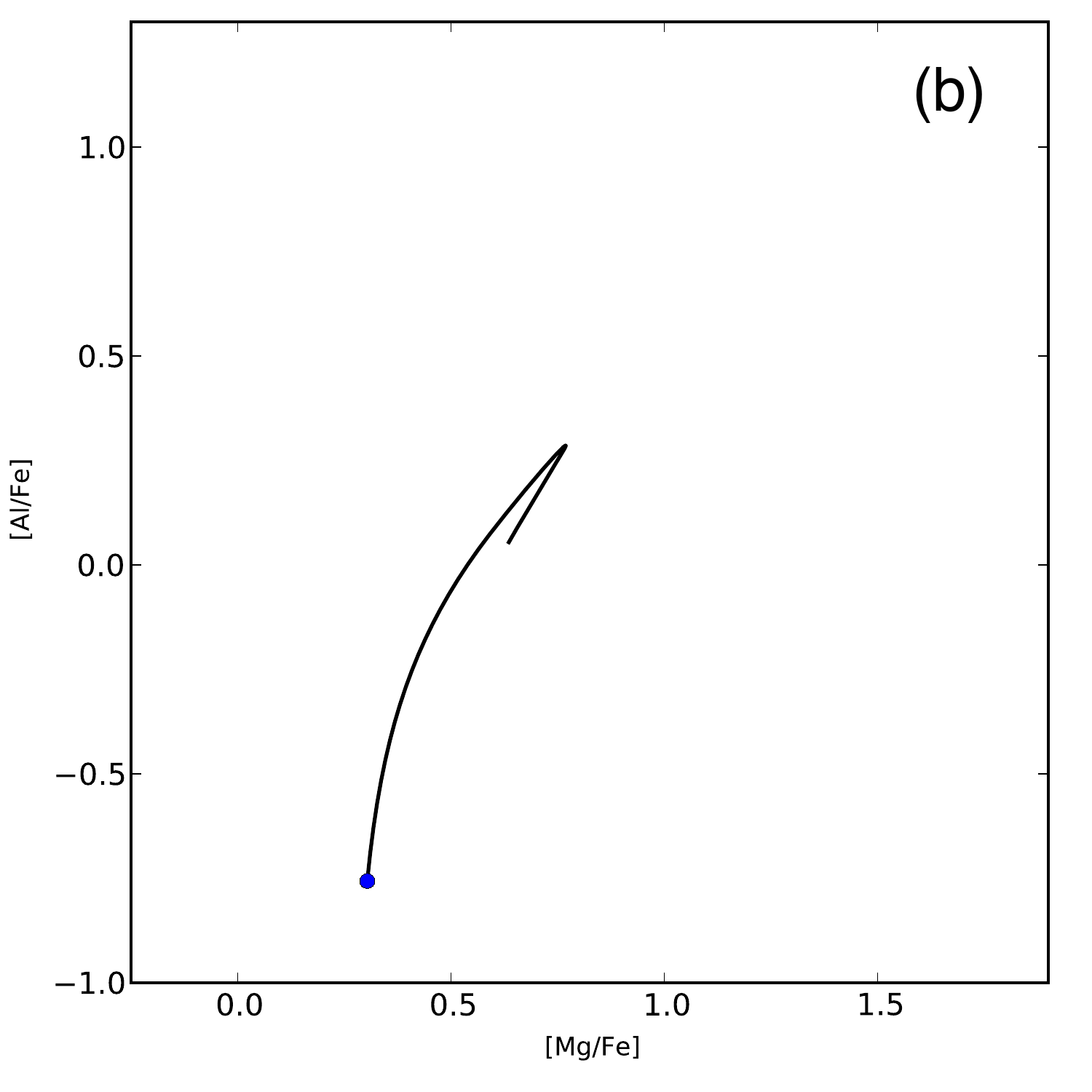}
 \end{center}
 \caption{Chemical evolution results of Na, O, Al, and Mg with Evel ChemEvol for comparison with \citet[][Figure 1]{Fenner:2004ju}. The blue point indicates the composition after the massive star pollution phase and before ejecta from AGB stars has been produced.}
 \label{fig:fennertestnaomg}
\end{figure}

\begin{figure}
 \begin{center}\includegraphics[width=0.5\columnwidth]{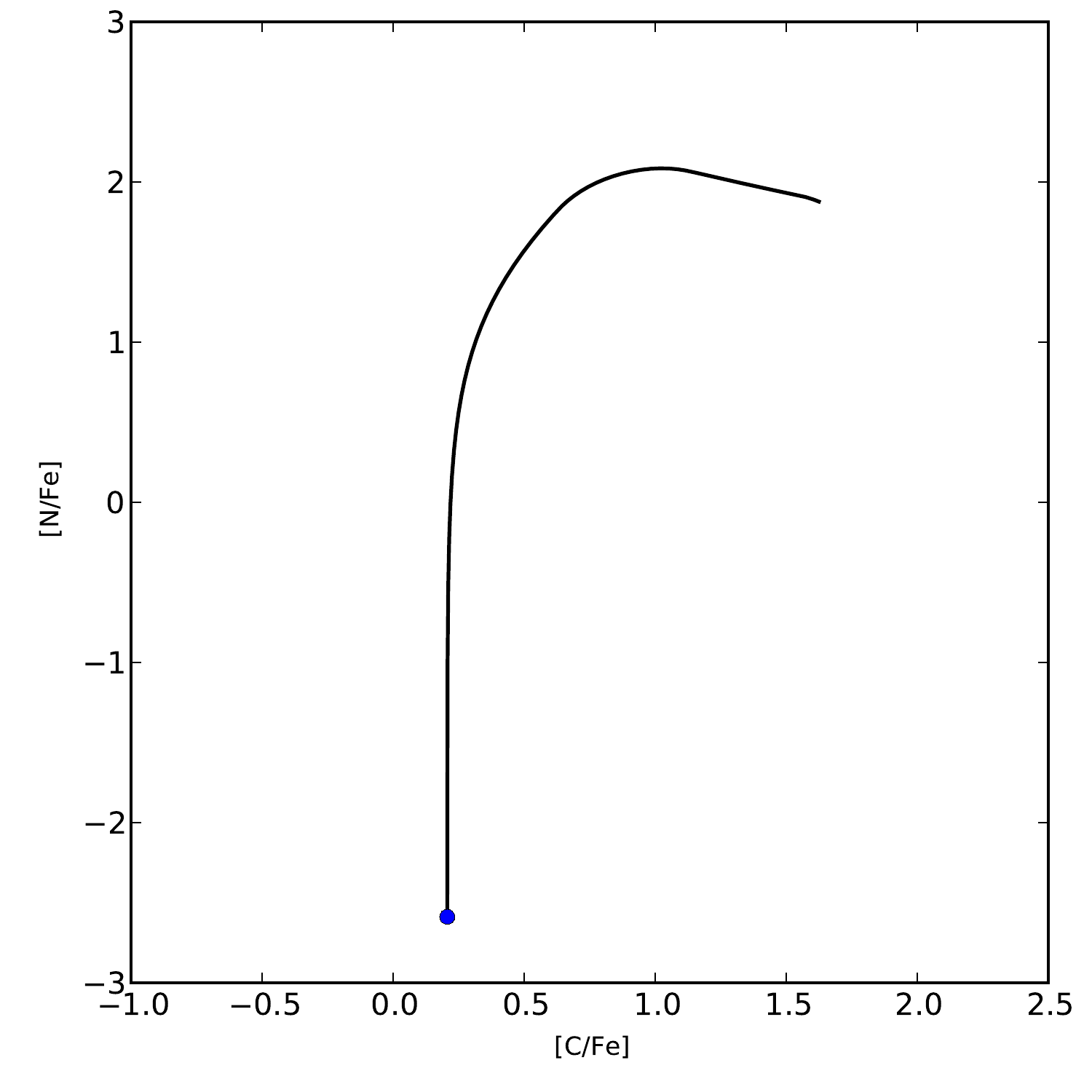}\end{center}
 \caption{Chemical evolution results of N and C with Evel ChemEvol for comparison with \citet[][Figure 3]{Fenner:2004ju}.}
 \label{fig:fennertestnc}
\end{figure}

\begin{figure}
 \begin{center}\includegraphics[width=0.5\columnwidth]{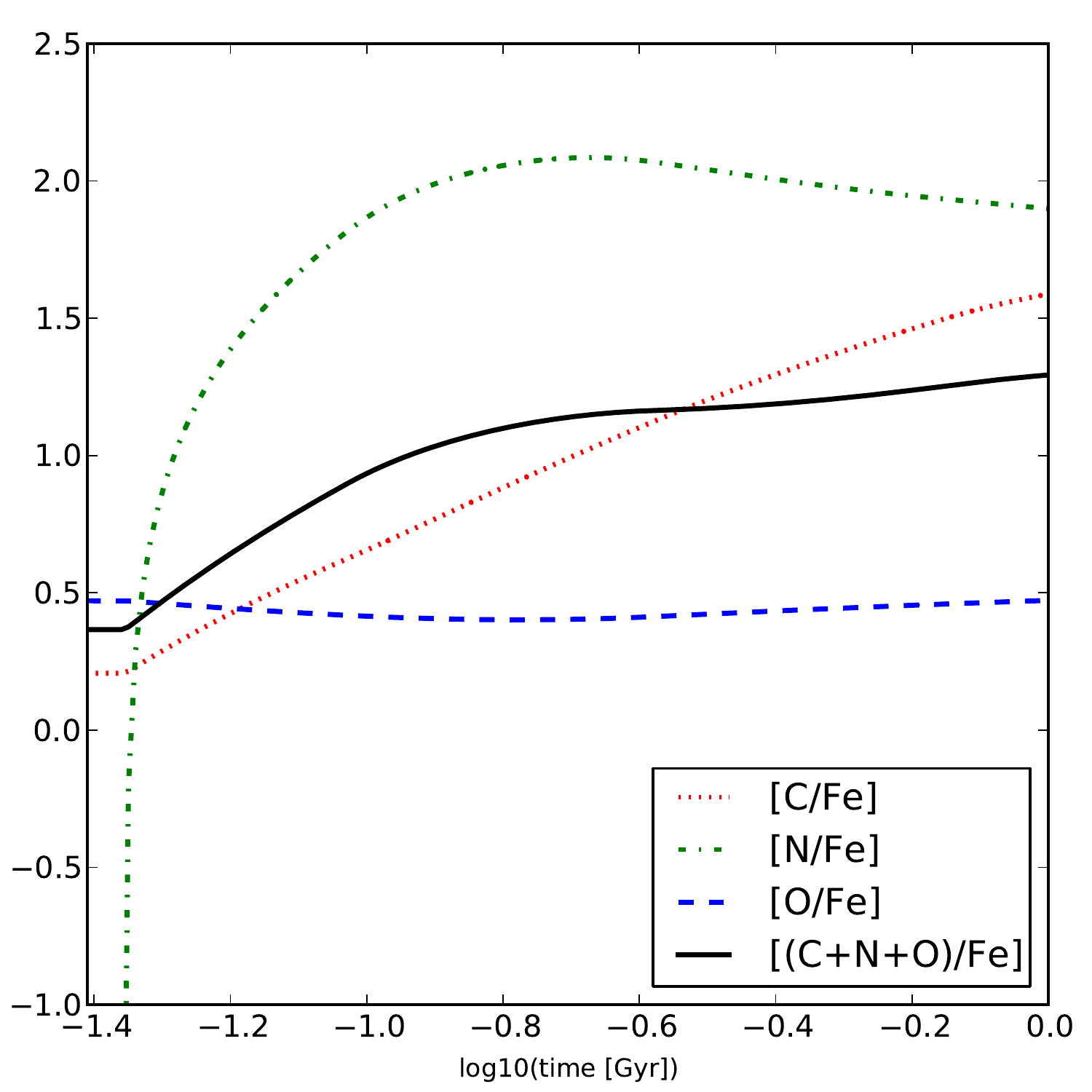}\end{center}
 \caption{Chemical evolution results of C, N, and O with Evel ChemEvol for comparison with \citet[][Figure 4]{Fenner:2004ju}.}
 \label{fig:fennertestcno}
\end{figure}

Our chemical evolution results in Figures \ref{fig:fennertestnaomg}, \ref{fig:fennertestnc}, and \ref{fig:fennertestcno} for Evel ChemEvol correspond almost exactly to Figure 1, 3, and 4 of \citet{Fenner:2004ju}. Small differences in the output can be explained by differences in the stellar lifetime function and the treatment of the massive star pollution phase, which were not specified in detail in the original paper.

The results of this comparison give us confidence that the Evel ChemEvol is producing the correct abundance outputs and can be used to explore new chemical evolution scenarios.

\end{document}